\documentclass[a4paper,9pt, onecolumn]{article}
\usepackage{lmodern}
\usepackage[utf8]{inputenc}
\usepackage[T1]{fontenc}
\usepackage[english]{babel}
\usepackage[usenames,svgnames]{xcolor}
\usepackage{setspace}
\usepackage{endnotes}
\usepackage[labelfont=bf]{caption}
\usepackage{subcaption}
\usepackage{authblk}
\usepackage[pdftex]{graphicx} 
\usepackage[pdftex,pageanchor=true,hyperindex=true]{hyperref}
\usepackage{epstopdf} % to change .eps to .pdf
\usepackage{bookmark}
\usepackage{pdfpages}
\usepackage{array}
\usepackage{type1cm,soul}
\usepackage{tabularx}
\usepackage{ifthen} 
\usepackage{lscape}
\usepackage{amsmath}
\usepackage{amsfonts}
\usepackage{wasysym}
\usepackage{fullpage}
\usepackage{xstring}
\usepackage{titlesec}
\usepackage{colortbl}
\usepackage{tcolorbox}
\usepackage{calc}
\usepackage{enumitem}
\usepackage{gensymb} %for the degree symbol
\usepackage{pifont} %for checkmarks
\usepackage{booktabs} %for the tables
\usepackage{multirow} %for the tables
\usepackage{bm} %for bold math
\usepackage{siunitx}

\graphicspath{{Figures/}}

\title{A probabilistic deep learning model of inter-fraction anatomical variations in radiotherapy}

\author[1,2]{Oscar Pastor-Serrano}
\author[3,4]{Steven Habraken}
\author[3,4]{Mischa Hoogeman}
\author[1]{Danny Lathouwers}
\author[1,4]{Dennis Schaart}
\author[2]{Yusuke Nomura}
\author[2]{Lei Xing}
\author[1]{Zoltán Perkó}

\affil[1]{\small Delft University of Technology, Department of Radiation Science \& Technology, Delft, Netherlands}

\affil[2]{\small Stanford University, Department of Radiation Oncology, Stanford, CA, USA}

\affil[3]{\small Erasmus University Medical Center, Department of Radiotherapy, Rotterdam, Netherlands}

\affil[4]{\small HollandPTC, Department of Medical Physics and Informatics, Delft, Netherlands}

\date{}

\begin{document}
\newcommand{\FixRef}[3][sec:]
{\IfBeginWith{#2}{#3}
	{\StrBehind{#2}{#3}[\RefResult]}
	{\def\RefResult{#2}}\IfBeginWith{#1}{#3}
	{\StrBehind{#1}{#3}[\RefResultb]}
	{\def\RefResultb{#1}}}

\newcommand{\secref}[1]
{\FixRef{#1}{sec:}Section~\ref{sec:\RefResult}}
\newcommand{\secreff}[1]
{\FixRef{#1}{sec:}in Section~\ref{sec:\RefResult}}
\newcommand{\Secreff}[1]
{\FixRef{#1}{sec:}In Section~\ref{sec:\RefResult}}
\newcommand{\secrefm}[2]
{\FixRef[#2]{#1}{sec:}Sections~\ref{sec:\RefResult}-\ref{sec:\RefResultb}}
\newcommand{\secreffm}[2]
{\FixRef[#2]{#1}{sec:}in Sections~\ref{sec:\RefResult}-\ref{sec:\RefResultb}}
\newcommand{\Secreffm}[2]
{\FixRef[#2]{#1}{sec:}In Sections~\ref{sec:\RefResult}-\ref{sec:\RefResultb}}
\newcommand{\figref}[1]
{\FixRef{#1}{fig:}Figure~\ref{fig:\RefResult}}
\newcommand{\figrefm}[2]
{\FixRef[#2]{#1}{fig:}Figures~\ref{fig:\RefResult}-\ref{fig:\RefResultb}}
\newcommand{\figreff}[1]
{\FixRef{#1}{fig:}in Figure~\ref{fig:\RefResult}}
\newcommand{\figreffm}[2
]{\FixRef[#2]{#1}{fig:}in Figures~\ref{fig:\RefResult}-\ref{fig:\RefResultb}}
\newcommand{\Figreff}[1]
{\FixRef{#1}{fig:}In Figure~\ref{fig:\RefResult}}
\newcommand{\Figreffm}[2]
{\FixRef[#2]{#1}{fig:}In Figures~\ref{fig:\RefResult}-\ref{fig:\RefResultb}}
\newcommand{\tabref}[1]
{\FixRef{#1}{tab:}Table~\ref{tab:\RefResult}}
\newcommand{\tabreff}[1]
{\FixRef{#1}{tab:}in Table~\ref{tab:\RefResult}}
\newcommand{\Tabreff}[1]
{\FixRef{#1}{tab:}In Table~\ref{tab:\RefResult}}
\newcommand{\tabrefm}[2]
{\FixRef[#2]{#1}{tab:}Tables~\ref{tab:\RefResult}-\ref{tab:\RefResultb}}
\newcommand{\tabreffm}[2]
{\FixRef[#2]{#1}{tab:}in Tables~\ref{tab:\RefResult}-\ref{tab:\RefResultb}}
\newcommand{\Tabreffm}[2]
{\FixRef[#2]{#1}{tab:}In Tables~\ref{tab:\RefResult}-\ref{tab:\RefResultb}}
\newcommand{\egyref}[1]
{\FixRef{#1}{eq:}Equation~\ref{eq:\RefResult}}
\newcommand{\eqreff}[1]
{\FixRef{#1}{eq:}in Equation~\ref{eq:\RefResult}}
\newcommand{\Eqreff}[1]
{\FixRef{#1}{eq:}In Equation~\ref{eq:\RefResult}}
\newcommand{\eqrefm}[2]
{Equations~\ref{eq:#1}-\ref{eq:#2}}
\newcommand{\eqreffm}[2]
{\FixRef[#2]{#1}{eq:}in Equations~\ref{eq:\RefResult}-\ref{eq:\RefResultb}}
\newcommand{\Eqreffm}[2]
{\FixRef[#2]{#1}{eq:}In Equations~\ref{eq:\RefResult}-\ref{eq:\RefResultb}}
\newcommand{\charef}[1]
{\FixRef{#1}{cha:}Chapter~\ref{cha:\RefResult}}
\newcommand{\chareff}[1]
{\FixRef{#1}{cha:}in Chapter~\ref{cha:\RefResult}}
\newcommand{\Chareff}[1]
{\FixRef{#1}{cha:}In Chapter~\ref{cha:\RefResult}}

\newcommand*{\dd}{\mathrm{d}}
\newcommand{\ui}[1]{\textit{\textbf{#1}}}
\newcommand{\mx}[1]{\underline{\underline{#1}}}
\newcommand{\diff}[2]{\dfrac{\dd #1}{\dd #2}}
\newcommand{\pdiff}[2]{\dfrac{\partial #1}{\partial #2}}
\newcommand{\dhl}{\hline\hline}
\newcommand{\rb}[1]{\left(#1\right)}
\newcommand{\sqb}[1]{\left[#1\right]}
\newcommand{\tb}[1]{\left<#1\right>}
\newcommand{\cb}[1]{\left\{#1\right\}}
\newcommand{\abs}[1]{\left|#1\right|}
\newcommand{\dspm}[1]{\begin{displaymath}#1\end{displaymath}}
\newcommand{\ds}{\displaystyle}
\newcommand{\pow}[2]{\cdot #1^{#2}}
\newcommand{\evat}[2]{\left.#1\right|_{#2}}
\newcommand{\ifrac}[2]{\ds #1 / #2}
\newcommand{\ab}{\ifrac{\alpha}{\beta}}
\newcommand{\BED}{\text{BED}}
\newcommand{\norm}[1]{\left\lVert#1\right\rVert}

\newcommand{\cmark}{\ding{51}}%
\newcommand{\xmark}{\ding{55}}%

\def\thetable{\Roman{table}}
\thispagestyle{empty}
\onecolumn
\maketitle
\noindent

\begin{abstract}
Objective: In radiotherapy, the internal movement of organs between treatment sessions – unless accounted for by time-consuming adaptation – causes errors in the final radiation dose delivery. To assess the need for adaptation, motion models can be used to simulate dominant motion patterns and assess anatomical robustness before delivery. Traditionally, such models are based on principal component analysis (PCA) and are either patient-specific (requiring several scans per patient) or population-based, applying the same set of deformations to all patients. We present a hybrid approach which, based on population data, allows to predict patient-specific inter-fraction variations for an individual patient.

Approach: We propose a deep learning probabilistic framework that combines patient-specific and population information to generate deformation vector fields (DVFs) warping a patient’s planning computed tomography (CT) into  possible patient-specific anatomies. This daily anatomy model (DAM) uses few random variables capturing groups of correlated movements. Given a new planning CT, DAM estimates the joint distribution over the variables, with each sample from the distribution corresponding to a different deformation. We train our model using dataset of 312 CT pairs with prostate, bladder, and rectum delineations from 38 prostate cancer patients. For 2 additional patients (22 CTs), we compute the contour overlap between real and generated images, and compare the sampled and “ground truth” distributions of volume and center of mass changes. 

Results: With a DICE score of $0.86\pm0.05$ and a distance between prostate contours of $1.09\pm0.93$ mm, DAM matches and improves upon previously published PCA-based models, using as few as 8 latent variables. The overlap between distributions further indicates that DAM's sampled movements match the range and frequency of clinically observed daily changes on repeat CTs.

Significance: Conditioned only on planning CT values and organ contours of a new patient without any pre-processing, DAM can accurately predict CTs and structures seen during following treatment sessions, which can be used for anatomically robust treatment planning and robustness evaluation against inter-fraction anatomical changes.
\end{abstract}

\section{Introduction}
\label{sec:Introduction}
% proton therapy, sensitivity to uncertainties, anatomical uncertainties
Modern radiotherapy techniques such as intensity modulated proton therapy (IMPT) have the potential to deliver highly conformal doses to tumors while maximally sparing organs at risk (OARs). Although offering dosimetric advantages with respect to conventional RT modalities, such treatments are particularly sensitive to geometrical uncertainties arising from setup errors before delivery or range errors caused by organ movements between or during treatment sessions. In the presence of uncertainties, planned doses are delivered to anatomies different from the 3D computed tomography (CT) scan used during treatment planning, which may translate into shifting high dose regions away from clinical target volumes (CTVs) into critical OARs. Being one of the main sources of error in, e.g., prostate cancer treatments \cite{van_herk_inclusion_2002}, the magnitude of the deformations and their effect on the final dose distribution must be quantified to ensure robust delivery. Ideally, treatments could be real-time adapted via image guidance, or alternatively adjusted before each treatment session \cite{jagt_automated_2018, jagt_near_2017}, but such adaptive workflows are constrained by the speed of the CT acquisition, delineation, dose calculation and treatment re-optimization processes in practice.  

% non dl models
An efficient alternative currently used in the clinic consists of including setup and range uncertainties during treatment planning optimization to design robust treatment plans that withstand positioning and range errors \cite{unkelbach_robust_2018, rojo-santiago_accurate_2021, van_der_voort_robustness_2016}. Similarly, inter-fractional movement information could be incorporated during treatment planning or treatment evaluation to make treatment plans robust against complex geometrical variations. To account for such anatomical changes, some published works propose computing expected dose distributions using weighted scenarios, where each scenario corresponds to the dose deposited in a patient geometry generated by an anatomy model. Typically, such models extract the main eigenmodes of organ deformation --- groups of correlated movements ---  via principal component analysis (PCA) \cite{sohn_modelling_2005, jeong_bilinear_2010, budiarto_population-based_2011, szeto_population_2017}. During the last decades, linear PCA models have been successfully employed to quantify and understand the effect of organ deformations in different treatment sites and modalities \cite{thornqvist_adaptive_2013, rios_population_2017, magallon-baro_modeling_2019}; to extend clinical volumes with extra margins and compensate for anatomical changes \cite{thornqvist_treatment_2013, bondar_statistical_2014}; to characterize respiratory deformations \cite{zhang_patient-specific_2007, badawi_optimizing_2010}; and to simulate dosimetric outcomes of delivery in the presence of geometrical uncertainties \cite{sohn_dosimetric_2012, nie_organ_2012, xu_coverage-based_2014, tilly_dose_2017}. Focusing on conventional photon-based RT modalities, most of these studies are based only on organ contours without including CT intensity values, and require time-consuming image registrations as pre-processing to find corresponding points across a population of patients before being usable for learning generic deformations. Furthermore, all previously introduced models are either patient-specific (requiring several CTs per patient) or population-based (applying the same set of deformations to all patients), which limits their accuracy and applicability. For widespread adoption of anatomically robust treatment planning, we require accurate probabilistic models quickly generating patient-specific treatment anatomies.

% dvf and registration. Models
All published PCA models learn correlated organ movements from a dataset of 3D deformation vector fields (DVFs), where each vector indicates the magnitude and direction of displacement for each point in a voxelized volume. Such DVFs can be obtained via image registration algorithms finding a non-linear correspondence between, e.g., two CT scans \cite{vasquez_osorio_novel_2009, ashburner_fast_2007, bruveris_geometry_2015}. While traditional not data-driven algorithms require  minutes to solve a registration task, recent deep learning based methods reduce computing times down to few seconds and additionally increase registration accuracy \cite{balakrishnan_voxelmorph_2019, de_vos_end--end_2017}, typically using 2D \cite{ronneberger_u-net_2015} or 3D \cite{cicek_3d_2016} U-net convolutional architectures in combination with spatial transformer networks \cite{jaderberg_spatial_2015}. Several architectures generating DVFs and warping pairs of images have been proposed and applied to radiotherapy problems such as 4D image registration of moving images due to breathing \cite{lei_4d-ct_2020, romaguera_prediction_2020} or automated contour propagation in adaptive workflows \cite{liang_automated_2021}. 

% DL models: unc segmentation, registration, combined DL with prob
Our objective, however, is to generate a set of DVFs to warp a single planning CT into different repeat CTs that are likely to be observed during the course of a radiotherapy treatment. Ideally, a suitable model would be able to implicitly capture the relative likelihood of correlated groups of movements depending on the input patient geometry. Probabilistic frameworks based on variational inference \cite{blei_variational_2017,kingma_auto-encoding_2014, rezende_stochastic_2014} have been successfully applied to model uncertainty in organ segmentation tasks \cite{baumgartner_phiseg_2019, kohl_probabilistic_2018, kohl_hierarchical_2019, hu_supervised_2019}, making use of auxiliary latent variables that represent the main factors of variation behind the model's predictions. Similar probabilistic U-net based architectures have also been proposed for pure image registration tasks \cite{dalca_unsupervised_2019,krebs_learning_2019}, with applications to unsupervised contouring problems \cite{dalca_unsupervised_2019-1} and breathing movement prediction based on motion surrogates \cite{romaguera_probabilistic_2021}.

% Dam
Extending on these recent architectures, we present a probabilistic deep learning framework that represents common anatomical movements and deformations in a population of patients using few latent variables. The proposed daily anatomy model (DAM) first generates DVFs conditioned on an input planning CT scan and latent variables, where each combination of latent variables corresponds to a different group of movements; and subsequently warps the planning CT with the generated DVFs into a set of artificial repeat scans. We train the model using a dataset containing planning and repeat CTs recorded at different stages of prostate cancer treatments in three different institutions, evaluating whether DAM is able to learn realistic movements with two external patients. Compared to previous methods, DAM does not require any pre-processing registration step and can in principle be applied to quickly simulate patient anatomies for treatment adaptation and robustness evaluation purposes.

\begin{figure}
    \centering
    \includegraphics[width=.95\textwidth]{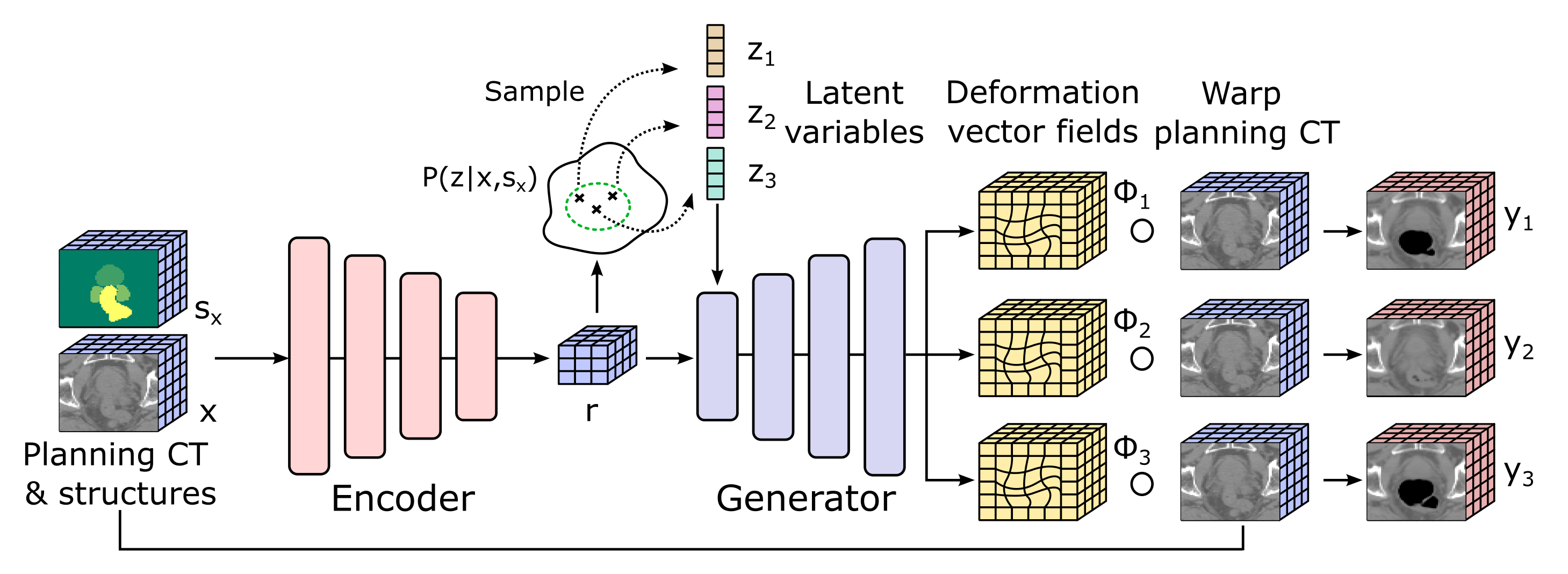}
    \caption{\textbf{Proposed generative framework.} The probabilistic models are embedded within a U-net, where the down-sampling path is referred to as Encoder, and the up-sampling path is the Generator. The Encoder takes the planning CT and structures and outputs both a compressed representation of the input $\bm{r}$ as well as a distribution $P(\bm{z}|\bm{x},\bm{s}_x)$ over the region of the latent space containing variables corresponding to plausible patient-specific movements. Given $\bm{r}$ and any sample $\bm{z}$ from the latent space distribution, the Generator outputs a deformation vector field that is used to warp the planning CT into an artificial repeat CTs.}
    \label{fig:problem}
\end{figure}

\section{Methods and materials}
\label{sec:Methods}
Here we describe the fundamentals of the variational framework used to capture anatomical variations, including the different parametric models and the procedure used to tune their parameters. Subsequently, we describe the model architecture in detail, together with the data and the evaluation metrics used in each experiment. 

\subsection{Proposed framework}
% Describe problem
% learning true distribution approximating data distribution
During the course of a radiotherapy treatment, the internal structures and organs of the patient change between fractions/days. As a result, the anatomy captured in the planning image $\bm{x}\in\mathbb{R}^{M}$ and organ structures $\bm{s}_x\in\mathbb{R}^{M}$ (both represented as 3D matrices) can significantly differ from the repeat images $\bm{y}\in\mathbb{R}^{M}$ and structures $\bm{s}_y\in\mathbb{R}^{M}$ taken during following treatment sessions. $M$ voxels comprise the entire volume, where the voxels in $\bm{x}$ and $\bm{y}$ represent image intensity values, and the voxels in $\bm{s}_x$ and $\bm{s}_y$ contain an integer corresponding to the organ present in the voxel.

As demonstrated in previous studies \cite{budiarto_population-based_2011} for treatment sites like prostate, common anatomical variations such as volume and contour changes are observed across an entire population. Based on the existence of such generic movements we assume that, given a planning image $\bm{x}$ and structures $\bm{s}_x$, there is an unknown patient-specific generative distribution $P ^*(\bm{y}|\bm{x},\bm{s}_x)$ of repeat scans that can be approximated via a probabilistic model with learned parameters. Given a planning image from a new patient, we can sample the resulting model distribution $P_{\bm{\theta}}(\bm{y}|\bm{x},\bm{s}_x)$ parametrized by $\bm{\theta}$ to generate a set of artificial anatomies observed at future treatment stages.

In this case, $\bm{\theta}$ corresponds to the parameters of the U-net neural network that is used to compute a DVF $\Phi:\mathbb{R}^3\rightarrow\mathbb{R}^3$ mapping coordinates between images. We model $\Phi$ as a diffeomorphic transformation, which is invertible, preserves topology, and in our case practically allows obtaining the forward and inverse transformations in a very simple manner. Such diffeomorphic transformation is represented by a stationary velocity field $v:\mathbb{R}^3\rightarrow\mathbb{R}^3$, as $\Phi=\exp{v}$. As for the inputs, we discretize the velocity field $\bm{v}\in\mathbb{R}^{M\times3}$ and DVF $\bm{\Phi}\in\mathbb{R}^{M\times3}$ into $M$ voxels, using $\bm{\Phi}(\bm{p})$ to denote the displacement applied to the voxel centered at location $\bm{p}\in\mathbb{R}^3$. Following previous work \cite{dalca_unsupervised_2019}, the U-net predicts $\bm{v}$, which is exponentiated via scaling and squaring using a spatial transformer network \cite{jaderberg_spatial_2015} (details in Appendix \ref{App:scsq}) to obtain the final DVF $\bm{\Phi}$ used to warp planning images into artificial repeats $\bm{y}=\bm{\Phi}\circ\bm{x}$.

\paragraph{Generative model}
% Describe model at testing time: sample prior and generate field
% describe latent variables and prior
We use a probabilistic model that conditions the generated DVFs (and thus also the repeat images) on $N$ unobserved latent variables $\mathbf{z}\in\mathbb{R}^N$ capturing the main factors of variation in the data, i.e., the main groups of anatomical deformations. The latent variables distribute following a multivariate Gaussian prior probability distribution that depends on the input planning anatomy

\begin{equation}
    P(\bm{z|\bm{x}},\bm{s}_x) = \mathcal{N}(\bm{\mu}_{\bm{\theta}}(\bm{x},\bm{s}_x), \bm{\Sigma}_{\bm{\theta}}(\bm{x},\bm{s}_x)),
\end{equation}

\noindent where the mean $\bm{\mu}_{\bm{\theta}}$ and diagonal covariance matrix $\bm{\Sigma}_{\bm{\theta}}$ are deterministic functions calculated by a neural network referred to as \textit{Encoder} (\figref{problem}), which corresponds to the down-sampling path of a U-net. The prior dependence on the input results in a different distribution over latent variables per patient, which allows the model to select the groups of movements that are likely to be observed for each specific input image. The Encoder additionally outputs a volume $\bm{r}=g_{\bm{\theta}}(\bm{x},\bm{s}_x)$, which is the results of several deterministic convolution operations containing features from the input. Since $\bm{r}$ is a deterministic function of the input, we substitute any conditioning on $\bm{r}$ with $\bm{x}$ and $\bm{s}_x$ in the remainder of the paper. 

The relationship between the input planning image and latent variables and the output warped repeat images is computed in the up-sampling path of the U-net, which takes sampled latent variables and the low-dimensional features $\bm{r}$ to generate a velocity field $\bm{v}_{\bm{z},\bm{\theta}}=f_{\bm{\theta}}(\bm{z},\bm{x},\bm{s}_x)$, where the subscripts denote the deterministic dependence to $\bm{z}$ and $\bm{\theta}$. After exponentiating $\bm{v}_{\bm{z},\bm{\theta}}$ to obtain the DVF $\bm{\Phi}_{\bm{z},\bm{\theta}}$, the output repeat image $\bm{y}\in\mathbb{R}^{M}$ is obtained by warping the input as $\bm{y} = \bm{\Phi}_{\bm{z},\bm{\theta}} \circ \bm{x}$.

Different latent variable samples $\bm{z}$ result in different repeat images given the same input planning scan, and the modeled distribution of repeat images can be recovered as a function of the prior $P(\bm{z|\bm{x}},\bm{s}_x)$ and a likelihood $P_{\bm{\theta}}(\bm{y}|\bm{z},\bm{x},\bm{s}_x)$ distributions as

\begin{equation}
	P_{\bm{\theta}}(\bm{y}|\bm{x},\bm{s}_x) = \int P_{\bm{\theta}}(\bm{y}|\bm{z},\bm{x},\bm{s}_x)P(\bm{z}|\bm{x},\bm{s}_x)d\bm{z}.
	\label{eq:marg}
\end{equation}  

The choice of the likelihood distribution affects the final loss function.  Based on previous work \cite{krebs_learning_2019}, we model the likelihood distribution as a function of the \textit{cross-correlation} (CC) between predicted $\bm{y}$ and ground-truth $\bm{\hat{y}}$ images, scaled by a constant $\lambda$ as

\begin{equation}
    P_{\bm{\theta}}(\bm{y}|\bm{z},\bm{x},\bm{s}_x) \propto \exp{(\lambda\text{CC}(\hat{\bm{y}},\bm{y}= \bm{\Phi}_{\bm{z},\bm{\theta}}\circ\bm{x}))}.
\end{equation}

The CC has been empirically found to yield better similarity than other metrics such as the mean squared error \cite{balakrishnan_voxelmorph_2019}, with larger CC values corresponding to more alike images. Let $y(\bm{p})$ and $\hat{y}(\bm{p})$ denote the intensity values for each voxel at position $\bm{p}$ in the predicted and ground-truth images, respectively. If $w(\bm{p})$ and $\hat{w}(\bm{p})$ are images where each voxel is the local mean of the $n^3$ neighbouring voxels, e.g., $w(\bm{p})=\frac{1}{n^3}\sum_{j=1}^{n^3}y(\bm{p}_j))$ and $\hat{w}(\bm{p})=\frac{1}{n^3}\sum_{j=1}^{n^3}\hat{y}(\bm{p}_j))$, the CC is defined as

\begin{equation}
    \text{CC}(\hat{\bm{y}},\bm{y}) = \sum_{\bm{p}\in\Omega}\frac{\big[\sum_{i=1}^{n^3}(\hat{y}(\bm{p}_i)-
    \hat{w}(\bm{p})) (y(\bm{p}_i)-w(\bm{p})) \big]^2}{\big[\sum_{i=1}^{n^3}(\hat{y}(\bm{p}_i)-\hat{w}(\bm{p})) \big]\big[\sum_{i=1}^{n^3}y(\bm{p}_i)-\hat{w}(\bm{p})) \big]}.
    \label{eq:cc}
\end{equation}

As in previous work \cite{krebs_learning_2019}, instead of sampling the likelihood $P_{\bm{\theta}}(\bm{y}|\bm{z},\bm{x},\bm{s}_x)$ each time during inference to generate anatomies, we always use the mode of the distribution $\bm{\Phi}_{\bm{z},\bm{\theta}}\circ\bm{x}$.

\paragraph{Learning} With the presented probabilistic formulation, the goal is to maximize \egyref{marg} by learning the parameters $\bm{\theta}$ from a dataset containing planning $\bm{x}$ and repeat $\bm{y}$ pairs. However, estimating the integral over the latent space would require sampling a large number of latent variables, being intractable in practice. Instead, we resort to a variational framework and define an \textit{approximate posterior} distribution $Q_{\bm{\psi}}(\bm{z}|\bm{x},\bm{s}_x,\bm{y},\bm{s}_y)$, parametrized by an \textit{Inference Network} with parameters $\bm{\psi}$. During training, the Inference Network has access to the real repeat scans and predicts the parameters of Gaussian distribution covering a small region of the latent space containing variables that are likely to explain the deformation between $\bm{x}$ and $\bm{y}$ scans. Thus, the predicted Gaussian is

\begin{equation}
    Q_{\bm{\psi}}(\bm{z|\bm{x},\bm{s}_x,\bm{y},\bm{s}_y}) = \mathcal{N}(\bm{\mu}_{\bm{\psi}}(\bm{x},\bm{s}_x,\bm{y},\bm{s}_y), \bm{\Sigma}_{\bm{\psi}}(\bm{x},\bm{s}_x,\bm{y},\bm{s}_y)),
\end{equation}

\noindent with deterministic mappings $\bm{\mu}_{\bm{\psi}}$ and $\bm{\Sigma}_{\bm{\psi}}$ computed by the Inference neural network. Our formulation allows estimating the model parameters $\bm{\theta}$ and $\bm{\psi}$ by minimizing the negative \textit{evidence lower bound} as

\begin{equation}
\log\:(P_{\bm{\theta}}(\bm{y}|\bm{x},\bm{s}_x)) \leq-\mathbb{E}_{\bm{z}\sim Q_{\bm{\psi}}(\bm{z}|\bm{x},\bm{s}_x,\bm{y},\bm{s}_y)}[\log(P_{\bm{\theta}}(\bm{y}|\bm{z},\bm{x},\bm{s}_x))] + D_{KL}(Q_{\bm{\psi}}(\bm{z}|\bm{x},\bm{s}_x,\bm{y},\bm{s}_y)||P_{\bm{\theta}}(\bm{z}|\bm{x},\bm{s}_x)).
\label{eq:elbo}
\end{equation}

The lower bound balances two terms: the $D_{KL}(\cdot ||\cdot)$ term --- Kullback - Leibler (KL) divergence --- forces the approximated posterior to be close to the prior distribution, while the first term corresponds to maximizing the CC, encouraging similarity between real and generated images. Further details about deriving the lower bound are included in Appendix \ref{App:elbo}. 

\paragraph{Explicit regularization terms} The current form of the likelihood enforces image similarity regardless of structure overlap or DVF quality. We modify the lower bound and add two regularization terms to enforce realistic predicted anatomies. To encourage smooth and realistic DVFs, we introduce a \textit{spatial} regularization term that penalizes large unrealistic spatial gradients $\nabla\bm{\Phi}_{\bm{z},\bm{\theta}}(\bm{p}) = \Big(\frac{\partial\bm{\Phi}_{\bm{z},\bm{\theta}}(\bm{p})}{\partial x}, \frac{\partial\bm{\Phi}_{\bm{z},\bm{\theta}}(\bm{p})}{\partial y}, \frac{\partial\bm{\Phi}_{\bm{z},\bm{\theta}}(\bm{p})}{\partial z} \Big)$ of the DVF $\bm{\Phi}_{\bm{z},\bm{\theta}}$, which is multiplied by a constant $\kappa$ as

\begin{equation}
    R(\bm{\Phi}_{\bm{z},\bm{\theta}}) = - \kappa\sum_{\bm{p}\in\Omega}\norm{\nabla\bm{\Phi}_{\bm{z},\bm{\theta}}(\bm{p})}_2
\end{equation}

A \textit{segmentation} regularization term is added to improve the overlap between propagated and ground-truth structures, using the DICE score (defined between 0 and 1, where 1 denotes perfect overlap). For $K$ structures, let $\bm{\hat{s}}_y^k$ be the voxels in the ground-truth scan with structure number $k\in[1,K]$, $\bm{s}_y^k = \bm{\Phi}_{\bm{z},\bm{\theta}}\circ\bm{s}_x^k$ the predicted voxels with structure number $k$, and $|\bm{\hat{s}}_y^k|$ the cardinality of structure $\bm{\hat{s}}_y^k$, i.e, the number of elements in $\bm{\hat{s}}_y^k$. The DICE score is defined as

\begin{equation}
    \text{DICE}(\bm{\hat{s}}_y^k, \bm{s}_y^k) = 2\frac{\abs{\bm{\hat{s}}_y^k\cap\bm{s}_y^k}}{\abs{\bm{\hat{s}}_y^k}+\abs{\bm{s}_y^k}}.
    \label{eq:dice}
\end{equation}

With these two terms multiplying the likelihood in the lower bound of \egyref{elbo}, the final optimization problem becomes

\begin{equation}
\begin{split}
\bm{\theta}^*,\bm{\psi}^* = \underset{\bm{\theta},\bm{\psi}}{\text{argmin}}\: \mathbb{E}_{\bm{x},\bm{y},\bm{s}_x,\bm{s}_y\sim P_{D}(\bm{x},\bm{y},\bm{s}_x,\bm{s}_y)}\Big[\mathbb{E}_{\bm{z}\sim Q_{\bm{\psi}}(\bm{z}|\bm{x},\bm{s}_x, \bm{y},\bm{s}_y)}\Big[\lambda\text{CC}(\hat{\bm{y}},\bm{y})-\frac{1}{K}\sum_{k=1}^K\text{DICE}(\bm{\hat{s}}_y^k, \bm{\Phi}_{\bm{z},\bm{\theta}}\circ\bm{s}_x^k))\\+\kappa\sum_{\bm{p}\in\Omega}\norm{\nabla\bm{\Phi}_{\bm{z},\bm{\theta}}(\bm{p})}_2\Big] + D_{KL}(Q_{\bm{\psi}}(\bm{z}|\bm{x},\bm{s}_x, \bm{y},\bm{s}_y)||P_{\bm{\theta}}(\bm{z}|\bm{x},\bm{s}_x))\Big],
\end{split}
\label{eq:elbo}
\end{equation}

 \noindent with $\bm{x}$, $\bm{y}$, $\bm{s_x}$ and $\bm{s}_y$ sampled from the real data distribution $P_{D}(\bm{x},\bm{y},\bm{s}_x,\bm{s}_y)$.

\subsection{Dataset}
To learn the model parameters in a training stage, we use a dataset with 369 CTs from 40 prostate cancer patients, including prostate, seminal vesicles, bladder and rectum delineations with no overlap. For each of the patients, 3-11 repeat CTs were recorded at different points during their treatment at 3 different institutions: Erasmus University Medical Center (Rotterdam, Netherlands), Haukeland Medical Center (Bergen, Norway) and the Netherlands Cancer Institute (Amsterdam, Netherlands) \cite{xu_coverage-based_2014,deurloo_quantification_2005,sharma_dose_2012}. In total, 329 planning-repeat CT pairs are available, 312 of which are used for training and validation, while the remaining 22 CTs --- corresponding to 2 independent test patients, as in previous studies \cite{budiarto_population-based_2011} --- serve to evaluate performance on unseen geometries. After rigidly aligning each repeat to the planning CT, we crop the volumes to a region of $64\times64\times48$ voxels around the prostate with a voxel resolution of 2 mm, resulting in sub-volumes of $128\times128\times96$ that in all cases covers the prostate, seminal vesicles, rectum and a large portion the bladder. As a result, we obtain $\bm{x}\in\mathbb{R}^{64\times64\times48}$ and $\bm{y}\in\mathbb{R}^{64\times64\times48}$ with the original CT intensity values rescaled to the range [0,1], and $\bm{s}_x\in\mathbb{R}^{64\times64\times48}$ and $\bm{s}_y\in\mathbb{R}^{64\times64\times48}$ with categorical labels depending on the organ present in each voxel. Given the stochasticity in the density of the rectum fillings, we adhere to clinical practice and mask all voxels in the rectum and set their intensity to -1000 (vacuum). 

\subsection{Model architecture} 
\begin{figure}[t]
    \centering
    \includegraphics[width=.95\textwidth]{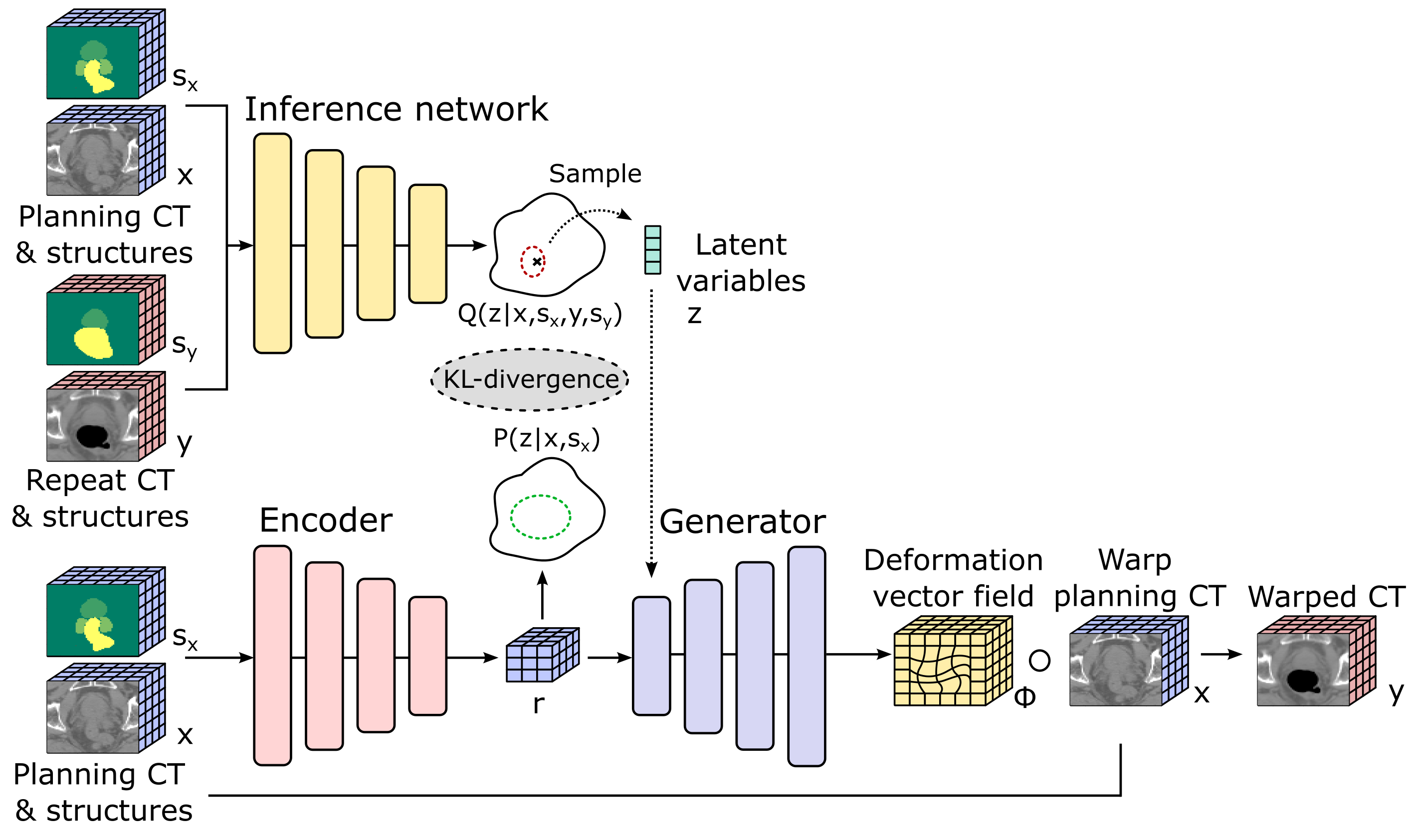}
    \caption{\textbf{Learning the model parameters}. An additional Inference Network takes a pair of planning and repeat CT and outputs the parameters of a distribution over a smaller region of the latent space that is likely to capture the deformation between the two images. The prior distribution predicted by the Encoder is forced to the distribution produced by the Inference Network via a KL-divergence term in the loss. Additionally, a reconstruction term encourages the resulting artificial CT (obtained after warping the planning scan with the predicted deformation) to be similar to real repeat CT.}
    \label{fig:arch}
\end{figure}

As shown in \figref{arch}, the proposed variational framework comprises two different models, parametrized by artificial neural networks: the Inference network and the probabilistic U-net with down-sampling and up-sampling paths denoted as Encoder and Generator, respectively. Based on the input planning CT and structures, the Encoder computes (i) a low-dimensional volume of input image features $\bm{r}$, and (ii) the parameters $\bm{\mu}_{\bm{\theta}}$ and $\bm{\Sigma}_{\bm{\theta}}$ of the prior distribution $P_{\bm{\theta}}(\bm{y}|\bm{z},\bm{x},\bm{s}_x)$ over a region of the latent space containing movements that are likely to be observed for the patient. The prior depends on the input, thus one of the functions of the Encoder is selecting primary groups of movements for each patient based on planning CT anatomy. The Generator takes the features $\bm{r}$ and sampled latent variables $z\sim P_{\bm{\theta}}(\bm{y}|\bm{z},\bm{x},\bm{s}_x)$ and produces the velocity field $\bm{v}_{\bm{z},\bm{\theta}}$ that is exponentiated to obtain a diffeomorphic transformation $\bm{\Phi}_{\bm{z},\bm{\theta}}$.

During training, the Inference network takes a pair of planning and repeat CTs and outputs the parameters $\bm{\mu}_{\bm{\psi}}$ and $\bm{\Sigma}_{\bm{\psi}}$ of the distribution $Q_{\bm{\psi}}(\bm{z}|\bm{x},\bm{s}_x,\bm{y},\bm{s}_y)$ over a much smaller region of the latent space containing latent variables that explain the deformation between both images. The DVF resulting from such latent variables is used to warp the planning CT into artificial repeat CTs $\bm{y}$ and structures $\bm{\Phi}\circ\bm{s}_x$. The distributions $Q_{\bm{\psi}}(\bm{z}|\bm{x},\bm{y})$ from the Inference network and $P_{\bm{\theta}}(\bm{z}|\bm{x})$ from the Encoder are forced to overlap via the KL divergence in \egyref{elbo}, while the artificial CT and structures are forced to match the ground-truth repeat CTs via the CC and DICE terms in the likelihood.

% Describe architecture
For the model with the lowest validation loss, the Encoder and Inference network are identical: three consecutive convolutional blocks, where each block contains a 3D convolutional layer with 32 channels and a $3\times3\times3$ kernel followed by Group Normalization \cite{wu_group_2020}, a rectified linear (ReLU) activation and a max pooling down-sampling operation. At the lowest level, an additional 3D convolution with 4 channels results in the low-dimensional feature volume $\bm{r}\in\mathbb{R}^{4\times8\times8\times6}$, which is mapped to the means and variances of the prior distribution via two different fully-connected layers. Conversely, the Generator first concatenates the latent variables to $\bm{r}$ as an additional channel, and then applies three up-sampling convolutional blocks with 32 channels. Two additional 3D convolution operations with 16 and 3 channels result in the final velocity field $\bm{v}_{\bm{z},\bm{\theta}}$. All models are trained for 1000 epochs using a learning rate of 0.001, hyper-parameters $\kappa=0.1$ and $\lambda=1000$, and the Adam optimizer \cite{kingma_adam_2017} with default parameters.

\subsection{Experiments}
We assess the model's accuracy in both generating feasible groups of deformations and reconstructing the ground-truth repeat scans. Additional experiments aim at exploring the structure of the latent space and the types of movements triggered by different latent variables.

\begin{itemize}
    \item Reconstruction accuracy. Given a planning and one of its repeat CTs in the test set, the Inference network can be used to obtain the latent variables corresponding to the deformation between both images, which are in turn used to get the DVF and warp the planning scan. For all 22 test planning/repeat pairs, we compare such generated repeat CTs to the ground truth repeats via computing the CC (\egyref{cc}) and the DICE score (\egyref{dice}). Additionally, we warp points 
    $\bm{\pi}_i\in\mathbb{R}^3$ on the surface of the planning prostate and calculate their distance to corresponding points $\hat{\bm{\pi}}_i\in\mathbb{R}^3$ on the surface of the repeat prostates via the mean surface error as
    
    \begin{equation}
        \bm{e}=\frac{1}{L}\sum_{i=1}^L\norm{\hat{\bm{\pi}}_i-\bm{\Phi}\circ\bm{\pi}_i}_2.
        \label{eq:surerr}
    \end{equation}

    To allow for a fair comparison with PCA-based methods, we compute the mean and standard deviation across the same $L=5864$ randomly chosen points as in previous studies \cite{budiarto_population-based_2011}. Finally, we evaluate the effect of the latent space dimensionality by comparing all accuracy metrics for different models trained with a varying number of latent variables.
    
    \item Generative performance. To finally be applied in clinical settings, the generated movements must match those from the recorded CT scans. Based on a previous study quantifying anatomical changes in prostate patients \cite{antolak_prostate_1998}, we compute the volume changes and center of mass shifts between planning and repeat scans, and compare their distributions obtained using real and artificial repeat CTs. To be able to compare to the reference values \cite{antolak_prostate_1998}, we reduce center of mass shifts to a single value by computing the average of absolute differences across coordinates.  
    
    \item Latent space analysis. By individually varying the values of each latent variable while keeping the other fixed, we numerically and visually assess the volume changes and center of mass shifts triggered by each variable. Finally, to understand the structure of the latent space, we obtain the latent variables from all pairs in the dataset and classify them according to the magnitude of their induced center of mass shifts and volume changes. Ideally, similar latent variables should correspond to similar deformations. which can be verified by plotting a 2D representation of the N latent variables using t-SNE \cite{maaten_visualizing_2008} together with their associated label to determine the presence of clusters. 
    
\end{itemize}

\section{Results}
\label{sec:Results}
In this section, we evaluate DAM's performance in generating realistic CTs with anatomical changes that match those of the real recorded repeat CTs. First, the reconstruction accuracy of real CTs is assessed, followed by an analysis of the latent space, and the types of deformations captured by the latent variables.

\subsection{Reconstruction accuracy}
\begin{figure}
    \centering
    \includegraphics[width=.99\textwidth]{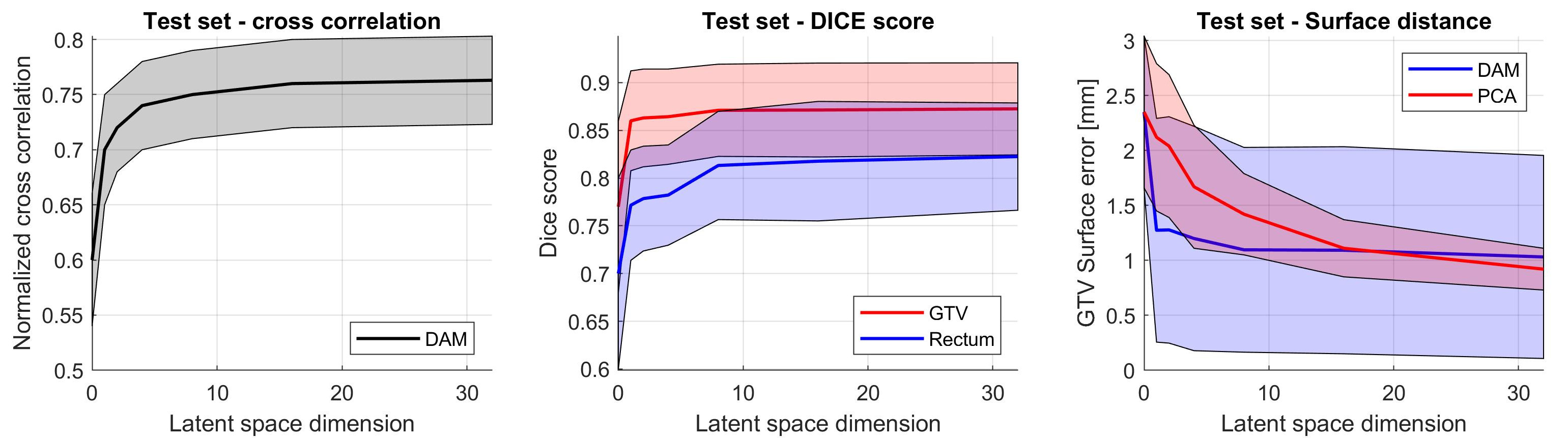}
    \caption{\textbf{Reconstruction accuracy metrics}. All figures show the mean (solid line) and standard deviation across all test planning-repeat pairs of the different metrics for a different number of latent variables, where 0 latent variables refers to using no model (always using the planning CT as a prediction). The left plot shows the cross-correlation between the real and reconstructed repeat CTs. In the middle plot, we show the DICE score measuring overlap between the warped planning structures and the organs delineated in the repeat CTs. Finally, the right figure shows the error between surface points in the prostate, compared to reference PCA values directly taken from \cite{budiarto_population-based_2011}.}
    \label{fig:rec}
\end{figure}

Given a planning-repeat pair of CT scans and structures in the test set, a repeat scan can be reconstructed via the same framework as used during training: sampling latent variables with the Inference network that are used by the Generator to generate a DVF. To verify the similarity between DAM's reconstructions and the real repeat CTs, we compute three metrics assessing CT and structure overlap: the CC, DICE score, and surface error $\bm{e}$. All three metrics in \figref{rec} are computed for different models trained with a varying number of latent variables, from 1 to 32. The values shown for 0 latent variables correspond to using the planning CTs as a prediction, which is equivalent to disregarding any model. First, the cross correlation between the real and reconstructed repeat CT is shown in the left plot of \figref{rec}, indicating that the model significantly improves when adding the first few variables, whereas no substantial is observed beyond 10 variables. As seen in DICE scores for the prostate and rectum from the middle plot in \figref{rec}, DAM can model prostate deformations with high accuracy even with a single latent variable, while representing rectum movements generally requires a slightly larger latent space with $\approx 8$ variables. The relative simplicity in capturing prostate movements is further confirmed from the right plot in \figref{rec}, showing that most surface error (\egyref{surerr}) reduction results from adding the first latent variable. On average, DAM matches --- and even outperforms in the low-dimensional regime --- the accuracy of countour-based PCA models \cite{budiarto_population-based_2011}. The larger spread in error values is likely caused by the fact that, unlike for the values reported in the PCA study, all surface points are not equidistant but randomly sampled over the surface, increasing the distance between correspondent points in under-sampled areas.

\subsection{Generative performance}
\begin{figure}
    \centering
    \begin{subfigure}[t]{0.47\textwidth}
        \centering
        \includegraphics[width=\textwidth]{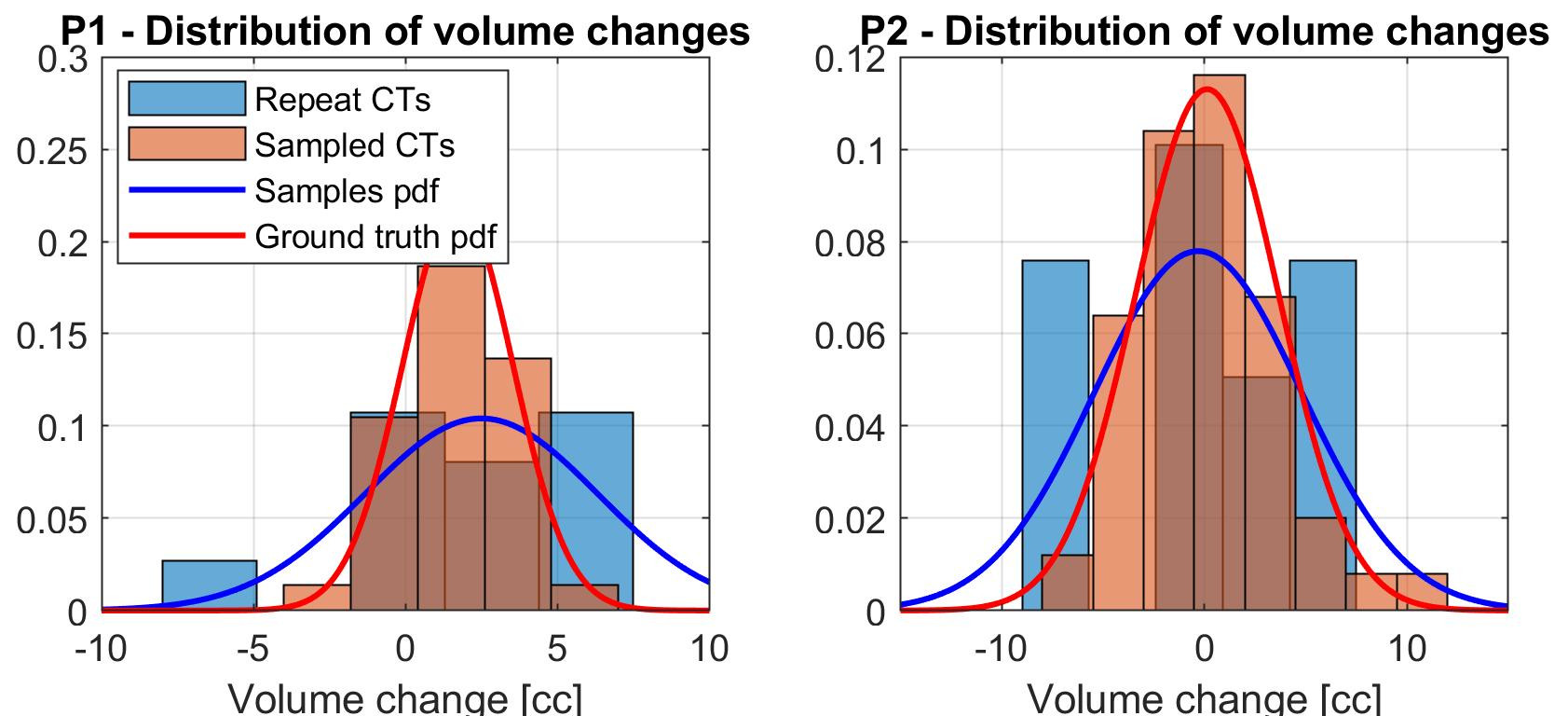}
        \caption{Volume changes}
        \label{fig:test_vol}
    \end{subfigure}
    \hfill
    \begin{subfigure}[t]{0.47\textwidth}
        \centering
        \includegraphics[width=\textwidth]{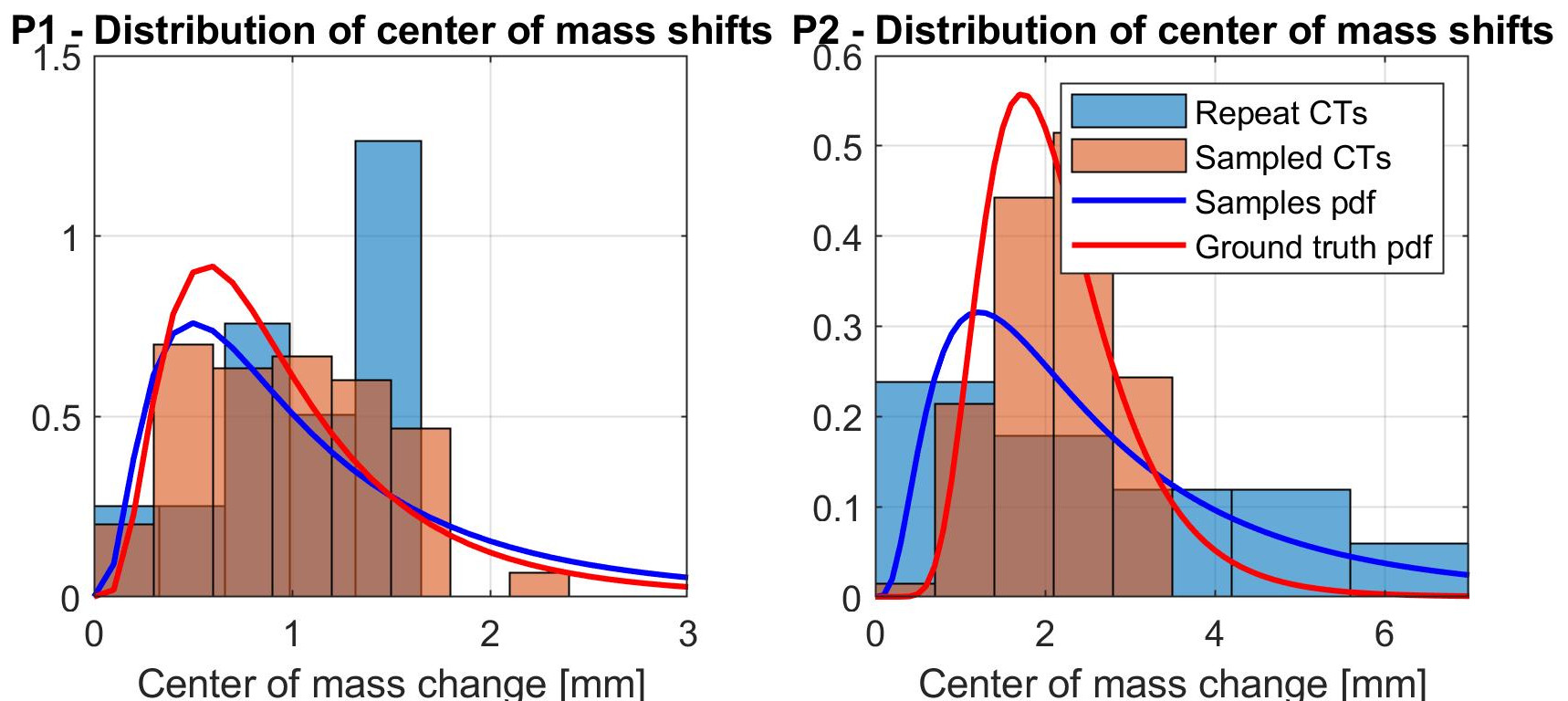}
        \caption{Center of mass shifts}
        \label{fig:test_com}
    \end{subfigure}
    \caption{\textbf{Test set histograms of anatomical variations}. For the two independent test patients, we plot histograms of prostate (a) volume changes and (b) center of mass shifts. Blue histograms correspond to changes between the planning CT and the 11 available repeat CTs, for which we additionally show their corresponding fitted normal and log-normal distributions in the same colors. Orange histograms are calculated using 100 randomly sampled CTs, obtained from 100 different latent variable combinations.}
    \label{fig:test_dist}
\end{figure}

\begin{figure}
    \centering
    \begin{subfigure}[t]{0.45\textwidth}
        \centering
        \includegraphics[width=\textwidth]{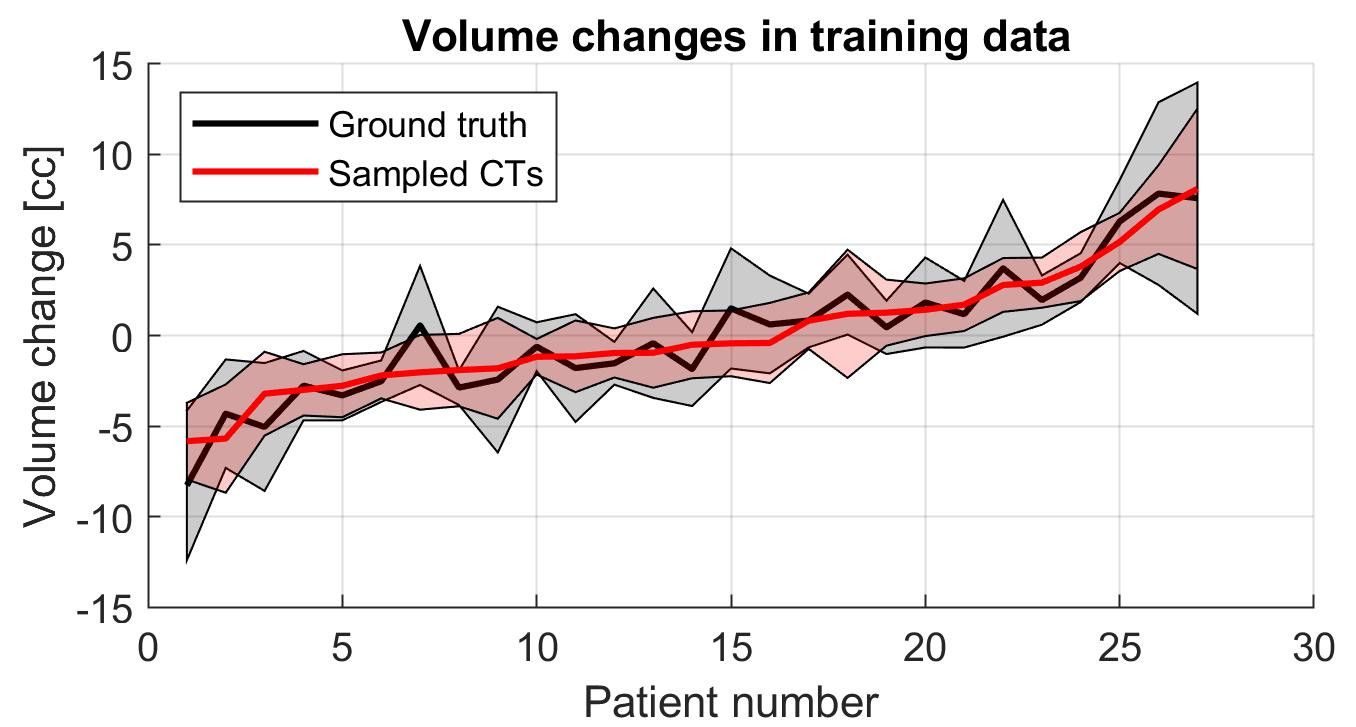}
        \caption{Volume changes}
        \label{fig:train_vol}
    \end{subfigure}
    \hfill
    \begin{subfigure}[t]{0.45\textwidth}
        \centering
        \includegraphics[width=\textwidth]{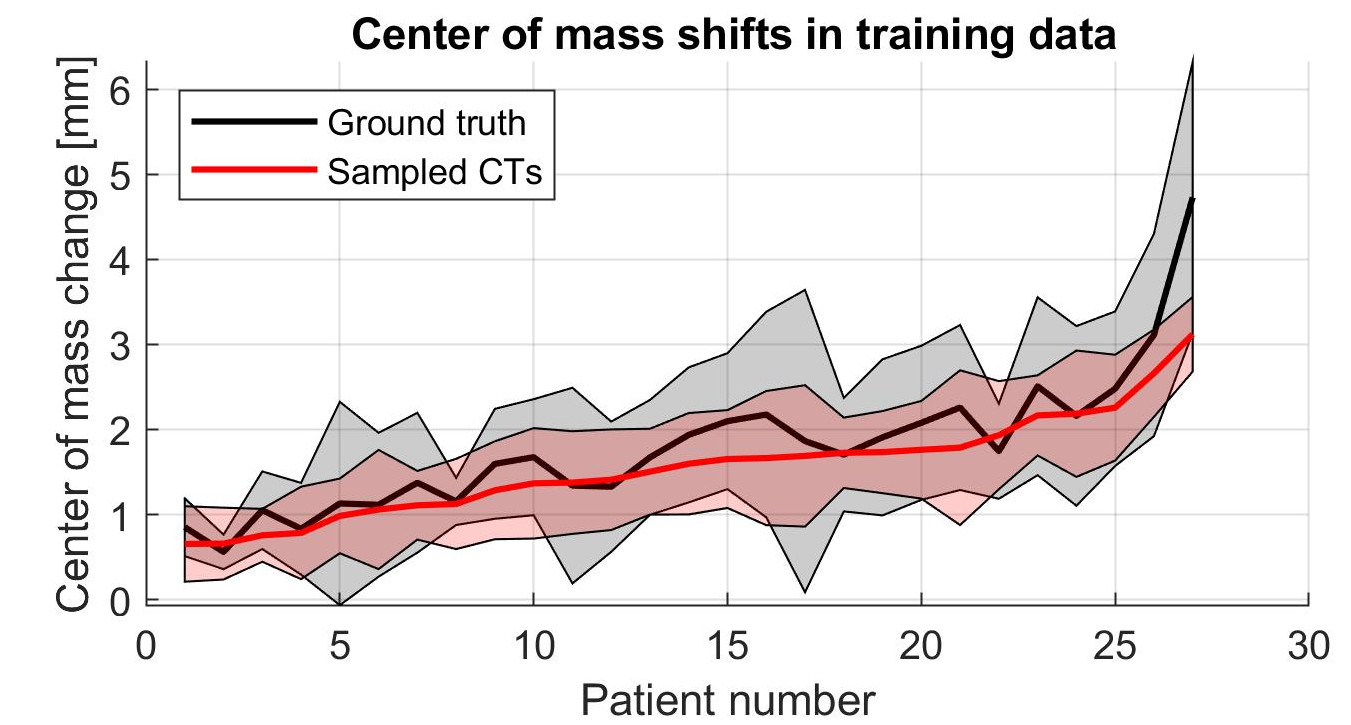}
        \caption{Center of mass shifts}
        \label{fig:train_Com}
    \end{subfigure}
    \caption{\textbf{Training set distribution of anatomical variations}. For all the patients in the training set with 5 or more repeat CTs, we plot the mean (solid line) and standard deviation of prostate (a) volume changes and (b) center of mass shifts. Black lines are computed using the available planning-repeat pairs of CT. The red curves are calculated using 100 randomly sampled CTs, obtained from 100 different latent variable combinations.}
    \label{fig:train_dist}
\end{figure}

\begin{figure}
    \centering
    \includegraphics[width=\textwidth]{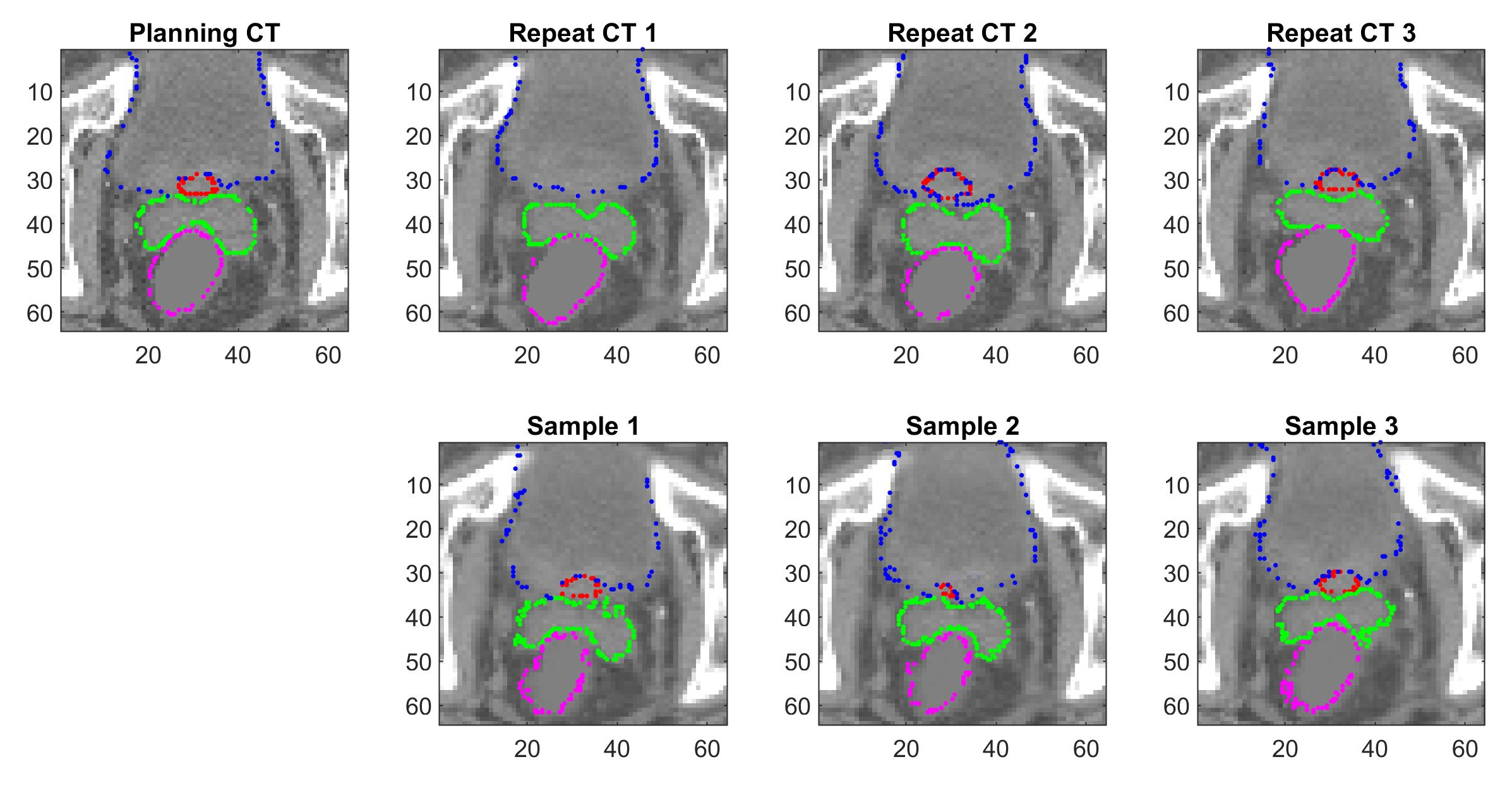}
    \caption{\textbf{Real vs. sampled anatomies}. Three recorded repeat CTs (top row), and three anatomies generated by the model (bottom row) are shown for one of the planning CTs, including prostate (red), seminal vesicles (green), bladder (blue) and rectum (pink) contours. The images correspond to a perpendicular slice in the cranial-caudal axis, showing the top of the prostate.}
    \label{fig:genan}
\end{figure}

Besides generating realistic CT scans, DAM should produce patient-specific movements whose distribution approximately matches those observed in the clinics, as reported in previous work \cite{antolak_prostate_1998}. For the 2 test patients, \figref{test_dist} displays the distribution of the anatomical variations seen in the 11 recorded repeat CTs (blue), compared to the deformations seen in 100 randomly sampled CTs (orange). Except for the large center of mass movements seen for the second patient \figref{test_com}, the ranges of values for both volume changes in \figref{test_vol}, and center of mass shifts in \figref{test_com} are approximately equal. Similarly, \figref{train_dist} shows the center of mass shift and volume changes distributions for all training patients with more than 5 repeat CTs. To compress all the information into one plot, we plot the mean and standard deviation, instead of the full histogram. The good overlap between distributions demonstrates that DAM captures the correct frequency and range of movements. As for the test patients, the biggest differences between both distributions occur for the last patient in \figref{fig:train_Com} with large center of mass shifts, which is aggravated by the fact that this patient has three big outliers of >7 mm shift. Finally, \figref{genan} displays generated and real anatomies for one of the patients, showing high quality images and contours with similar features and shapes.

\subsection{Latent space analysis}
\begin{figure}
    \centering
    \begin{subfigure}[b]{0.98\textwidth}
        \centering
        \includegraphics[width=\textwidth]{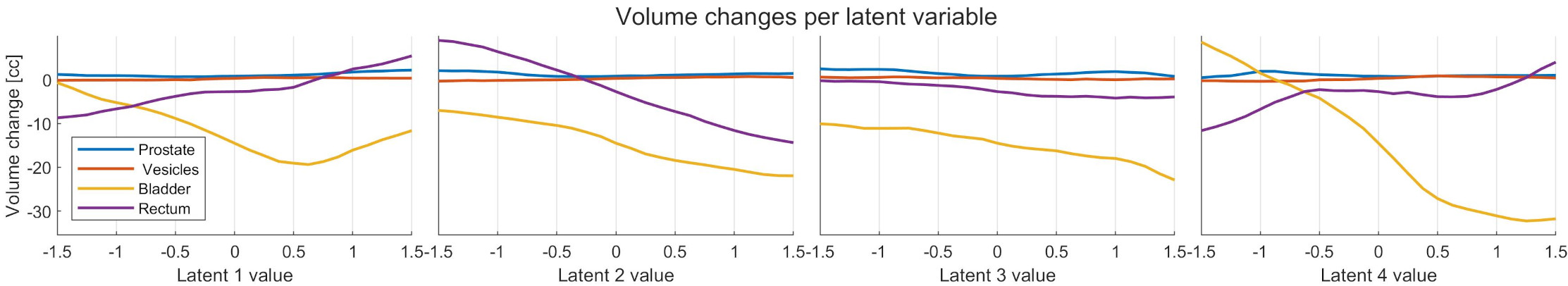}
        \caption{Volume changes}
        \label{fig:volch}
    \end{subfigure}
    
    \begin{subfigure}[b]{0.98\textwidth}
        \centering
        \includegraphics[width=\textwidth]{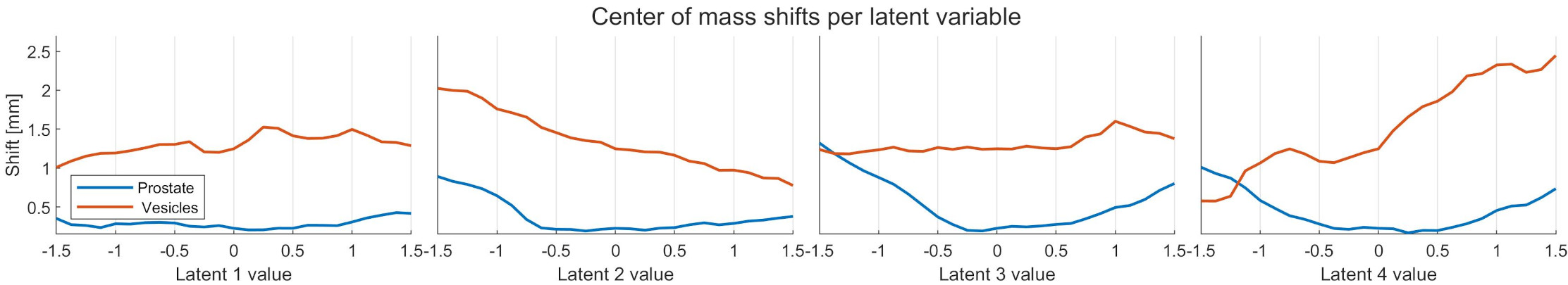}
        \caption{Center of mass shifts}
        \label{fig:comch}
    \end{subfigure}
    \caption{\textbf{Effect of individual latent variables on deformations}. (a) Volume changes and (b) center of mass shifts triggered by independently varying latent variables. For a model with $N=8$ latent variables, four randomly selected variables are varied between values within -1.5 and 1.5 of their standard deviation, while keeping the remaining seven variables fixed and equal to their mean.}
    \label{fig:changes}
\end{figure}

\begin{figure}
    \centering
    \includegraphics[width=.80\textwidth]{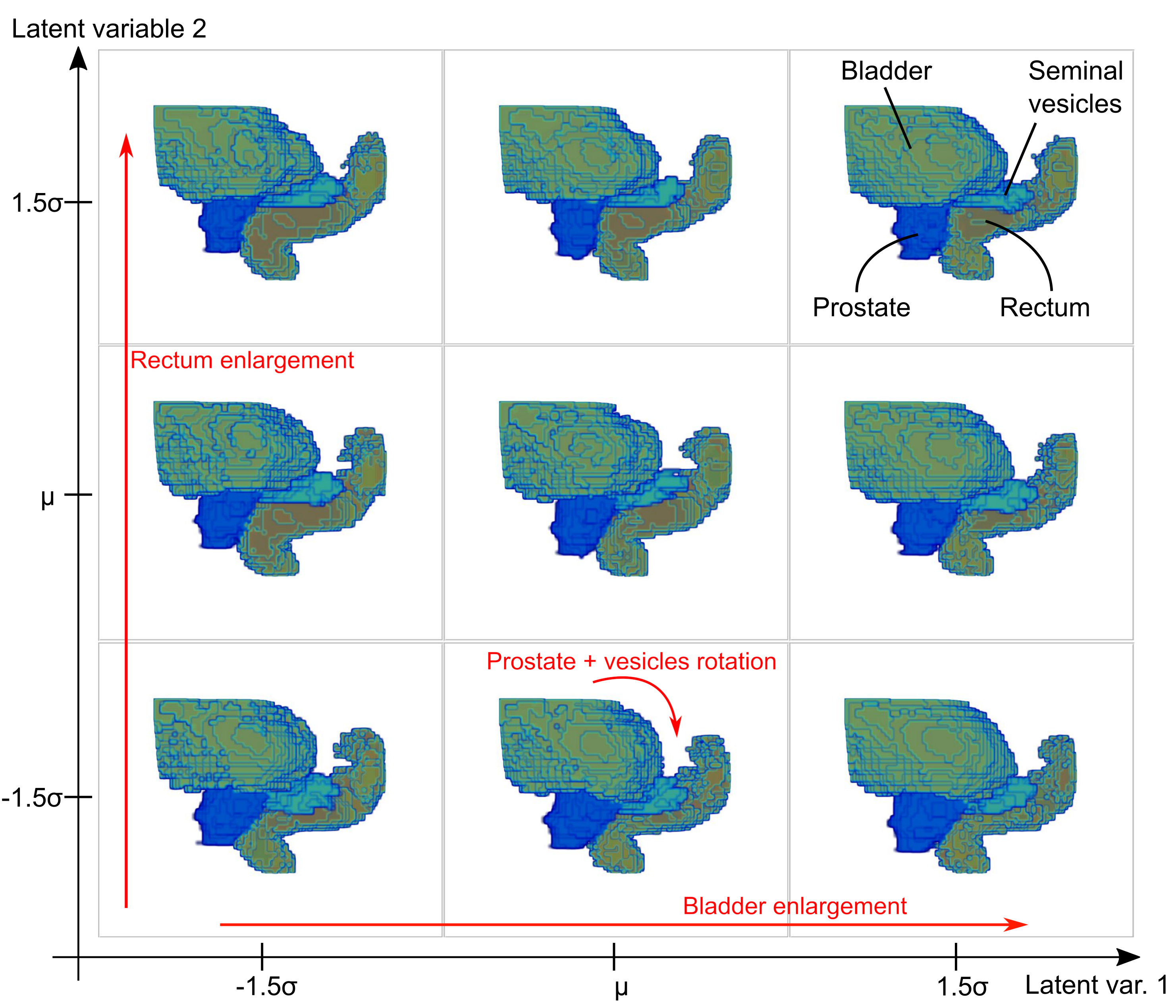}
    \caption{\textbf{Latent space visualization}. Grid plot of the prostate (blue), seminal vesicles (green), bladder (yellow) and rectum (orange) volumes. Each box corresponds to a different combination of latent variables in a 2D plane of the latent space, where the values for each variable are shown on the axes, with $\sigma$ being the standard deviation and $\mu$ the mean. Changes in the horizontal axis translate into bladder enlargements, while the vertical axis controls rectum volume. Correlated groups of movements are observed, e.g., as prostate rotations triggered by an enlarged bladder and smaller rectum.}
    \label{fig:grid}
\end{figure}

To investigate the deformations captured by the latent variables, we compute the center of mass shifts and volume changes triggered by each variable independently, while keeping the rest fixed. \figref{changes} displays such changes for 4 randomly picked variables from the model with 8 latent variables, whose value was modified between -1.5 and 1.5 times the standard deviation of the prior distribution. The results show magnitudes and correlations between changes as can be expected: smaller prostate volume changes, and large bladder and rectum variations shifting the center of mass of the prostate and seminal vesicles. To further demonstrate DAM's learned correlated groups of movements, in \figref{grid} we plot a grid of structures corresponding to simultaneously varying two latent variables. Individual changes in the horizontal and vertical axis mainly control the bladder and rectum volumes, respectively. Correlated deformations arise: the increase of bladder volume above the seminal vesicles, together with the decrease of rectum filling below the prostate, cause a prostate and vesicles shift and rotation.

\begin{figure}
    \centering
    \includegraphics[width=\textwidth]{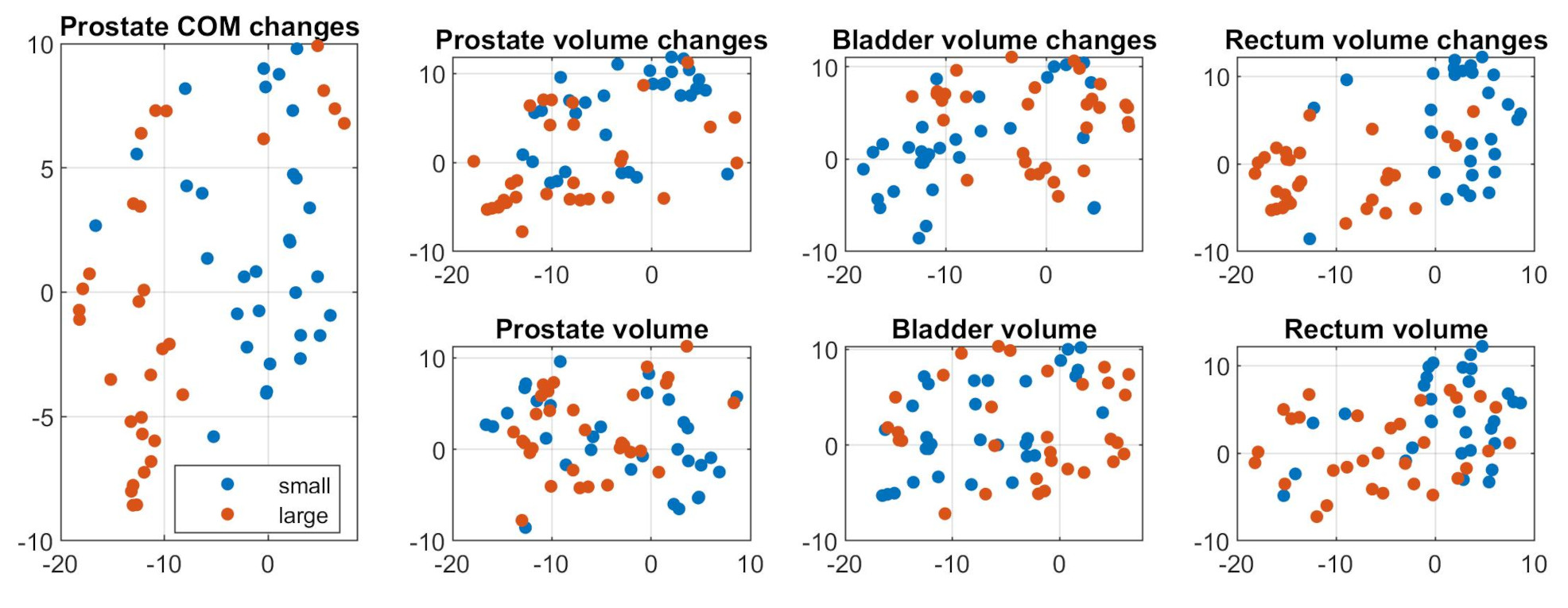}
    \caption{\textbf{Latent space structure}. Each latent variable is reduced to 2D space t-SNE representation and classified, from left to right, according to whether they correspond to small (blue) or large (orange): prostate center of mass shifts, prostate, bladder and rectum volume changes, or prostate, bladder and rectum sizes. 'Small' samples fall below the 10\% percentile of all values, while 'large' samples include all values above the 90\% percentile.}
    \label{fig:clusters}
\end{figure}

We analyze the structure of the latent space by determining if similar deformations (shifts and volume changes) or anatomical features (organ volume) result in similar latent variables. \figref{clusters} shows a two-dimensional t-SNE representation of the latent variables, where only samples with the smallest and largest movements or volumes are included, i.e., samples whose with center of mass shifts or volumes that fall above the 90\% percentile or below the 10\% percentile. Most of the latent space information seems to concern center of mass shifts and bladder/rectum volume changes, since their 2D representations can be clearly separated. Ideally, similar latent variables that are clustered together will correspond to different anatomical deformations, and will not carry information about anatomical features of the patient such as absolute organ volume. Instead, the Encoder is in charge to mapping deformations to anatomical traits observed in the planning CT or structures. Prostate and bladder volume seem to have no effect in how the latent space is organized, since similar latent variables correspond to very different sizes. To some extent, the effect of rectum size is also limited, resulting from the possible correlation between rectum fillings and volume changes.
% all bladders, regardless of size, will show large and small volume changes (sometimes full and sometimes empty and all intermediate)

%\begin{figure}
%    \centering
%    \begin{subfigure}[b]{0.98\textwidth}
%        \centering
%        \includegraphics[width=\textwidth]{train_latdist.jpg}
%        \caption{Training set patient}
%        \label{fig:testlatdist}
%    \end{subfigure}
%    
%    \begin{subfigure}[b]{0.98\textwidth}
%        \centering
%        \includegraphics[width=\textwidth]{test_latdist.jpg}
%        \caption{Test set patient}
%        \label{fig:trainlatdist}
%    \end{subfigure}
%    \caption{\textbf{Prior - posterior distribution overlap}. For (a) a training patient, and (b) a test patient, the prior Gaussian probability density function $P(\bm{z}|\bm{x}, \bm{s}_x)$ is compared to a normalized histogram of samples from the posterior distribution $Q(\bm{z}|\bm{x}, \bm{y}, \bm{s}_x, \bm{s}_y)$. The parameters of the prior distribution are obtained from the Encoder, given a planning CT and structure volume. Histograms are obtained by sampling once each posterior distribution corresponding to each of the planning-repeat pairs available for both patients.}
%    \label{fig:latdist}
%\end{figure}

\section{Discussion}
\label{sec:Discussion}
% side effect of regularizing the DVF
% effect of having few repeats
% Range matches literature
% struggle with larger deformations, not a problem. % Some deviations due to outliers or big deformations, for real use these are not taking into account in robust plans, but you rather adapt
In this study, we developed a probabilistic framework to model patient-specific inter-fraction movements based on population data. The presented DAM captures deformation patterns, generating DVFs only based on the planning CT scan and delineations. Based on the metrics obtained in \figref{rec} for the 22 scans from two independent test patients, DAM can generate realistic CTs with anatomical variations that resemble those recorded in the clinics using a small number of latent variables. The structure overlap of a model with a single variable, measured as a DICE score of $0.856\pm 0.058$, agrees with that of previous state-of-the-art pure segmentation/registration (non-generative) deep learning studies \cite{elmahdy_joint_2021,elmahdy_robust_2019,yuan_prostate_2019,liang_automated_2021}. Compared to linear PCA models where each eigenvector captures an independent mode of motion, the non-linearities in DAM allow representing different groups of correlated movements using different values of only one latent variable. Given that a single latent variable practically suffices to capture prostate movements, and that both the CC and rectum DICE score keep increasing with larger latent spaces, we can conclude that most of the computational effort is dedicated to modeling rectum deformations. Prostate IMPT treatments typically solely involve lateral beams, for which the impact of error due to rectum movement is small. In some cases, models with as little as 4-8 variables may be accurate enough, while 8-16 variables additionally ensure accurate rectum deformations for plans requiring more precision. 

For clinical application, it is critical that the model generates realistic shifts and deformations of the volume to be irradiated/treated (in this case, the prostate). Overall, based on the results in \figref{test_dist} and \figref{train_dist}, the center of mass shifts and volume changes produced by DAM show good overlap to the deformations and shifts recorded in the clinic, matching previously reported values \cite{antolak_prostate_1998}. One reason why DAM struggles in simulating the most extreme shifts or slides is the regularization term of the loss, which limits large deformations. Despite this limitation, such large anatomical variations are typically taken care of by adapting the treatment plan to the new anatomy, whereas robust treatment planning and evaluation --- the main potential applications of DAM --- are in principle oriented to incorporating average, frequent deformations into treatment design and evaluation, and we expect DAM to be useful for such purposes.

\paragraph{Comparison to other methods}
% PCA fails to generate, no check on whether sampling eigenmodes from the normal distribution results in realistic distribution of deformations
% Similar reconstruction, but on the entire CT. Trained using HU without needs of unifying pre-processing steps
% much less latent variables
All the previously published approaches are either patient-specific or population models based on PCA. Patient-specific methods \cite{sohn_modelling_2005, thornqvist_adaptive_2013, thornqvist_treatment_2013, nie_organ_2012,zhang_patient-specific_2007} require at least a few CTs recorded during a patient's treatment, and therefore they are unfeasible for pre-delivery robust treatment planning and evaluation, being restricted to post-treatment analysis. Conversely, population models \cite{budiarto_population-based_2011, magallon-baro_modeling_2019, tilly_dose_2017, rios_population_2017, szeto_population_2017} use a set of planning-repeat CT/contour pairs from previous patients, but simulate the same type of deformations for all patients regardless of their anatomy. In contrast, as seen in \figref{test_dist} and \figref{train_dist}, DAM is able to retrieve patient-specific magnitude and frequency of movements from the entire population based on the planning CT anatomy, making the model suitable for a wider range of applications.

Most previous studies \cite{budiarto_population-based_2011,magallon-baro_modeling_2019,sohn_modelling_2005,thornqvist_treatment_2013} model only the surface of the organs and not the intensities values in the CT. Without CT values the dose distributions are always calculated on the same planning CT with varying contours, which limits its applicability, especially in IMPT given the protons' finite range and tissue sensitivity.  Conversely, PCA-based models modeling full DVFs require 7 \cite{tilly_dose_2017} or up to 100 principal components \cite{szeto_population_2017} to capture 90\% of the variance in the training data. A large number of components (equivalent to DAM's latent variables) adds more variation, increases the chance of sampling unrealistic deformations and limits their applicability as reduced order models. Most importantly, all previous population-based methods require a time-consuming pre-processing step involving multiple deformable image registration steps between scans and patients to an organ or CT template. The accuracy of such registration calculation degrades the final accuracy and generative performance of the model, with previous studies \cite{szeto_population_2017,tilly_dose_2017} showing surface errors of around $1.5\pm1.0$ mm introduced in their pre-processing step alone that are comparable the DAM's total errors reported in the right plot of \figref{rec}. Given the lack of uniformity in treatment site and evaluation metrics in previous studies -- where most focus on evaluating the variance captured by the PCA model components and the errors on the DVFs caused by truncating the number of eigenmodes --- we compare DAM's performance to a PCA model of the prostate \cite{budiarto_population-based_2011} in the right plot of \figref{rec}. Even without adding any pre-processing errors, DAM matches the overall performance and is to capture prostate motion with a lower number of modeling parameters. Being trained directly on CT images in an unsupervised manner, DAM bypasses any performance or time losses from any pre-processing step, and can be easily applied to generate new anatomies in few milliseconds, compared to the tens of minutes or hours needed to obtain accurate enough registrations using conventional clinical software. 

Like PCA-based models, DAM assigns realistic correlated deformations to different values of the latent variables. \figref{changes} and \figref{grid} show that variables control shifts, volume changes and rotations similar to those reported in previous studies \cite{budiarto_population-based_2011, magallon-baro_modeling_2019}. \figref{clusters} demonstrates that the latent variables almost exclusively carry information about deformations, and not about anatomical traits from the patients. Instead, the Encoder is in charge of independently mapping planning anatomies to a subset of latent variables. Furthermore, unlike all previous approaches not evaluating the generative performance of their proposed models, we demonstrate the DAM also generates the adequate range and frequency of deformations for each patient.

\paragraph{Applicability} 
DAM's main application in robust treatment planning and robust evaluation involves sampling patient anatomies and calculating the corresponding dose distributions. With prediction times of few milliseconds per generated anatomy, DAM offers huge speed-up possibilities for plan evaluation when coupled to fast dose calculation algorithms \cite{pastor-serrano_millisecond_2022, pastor-serrano_learning_2022, perko_fast_2016, pastor-serrano_sub-second_2022}. Few (3-5) representative scenarios corresponding to points around mean of the posterior distribution can be sampled to be used for scenario based robust optimization, which may translate into a dosimetric advantage or be used for margin reduction. In principle, the same modeling framework could be applied to any other treatment site, with additional applications involving obtaining weighted dose scenarios \cite{tilly_dose_2017} to formulate anatomical robustness margin recipes \cite{van_der_voort_robustness_2016}.  Straightforward extensions include adding temporal dependence  for treatments where patients' anatomies significantly change following a clear pattern during (e.g., breathing) or between the different fractions of the treatment (e.g., modeling tumor shrinkage). Such time-dependent model could be coupled to breathing interplay effect simulation tools \cite{pastor-serrano_how_2021} to design plans based on breathing signals \cite{pastor-serrano_semi-supervised_2021} that mitigate the detrimental effect of movement during delivery.

\paragraph{Limitations}
% depends on the extent to which movements can be predicted only from anatomy
% rectum fillings
% effect of resolution and contrast in masking some small movements
% effect of dataset size 
Like PCA-based models, DAM will struggle to generate deformations that are not represented in the training data, for which continuously updating the model (e.g, using cone beam CTs) can be a solution. Likewise, low resolution images with poor contrast can also affect performance by masking small movements of structures, especially in areas with similar organ tissue densities. DAM's implementation in the clinic thus requires a quality assurance protocol that evaluates robustness in predictions e.g., by training several models using different data, and evaluating result similarity on a same test dataset. 

As for many other deep learning algorithms, DAM's generalization capabilities depend on the size and variability of the data in the dataset, as well as on the quality and resolution of the CT images. Due to the rather small size of the dataset in this preliminary study --- caused by the scarcity of recorded sets of planning and repeat CTs --- and based on the initial positive results, further testing appears warranted. 

DAM's accuracy in generating reasonable patient-specific movements depends on the extent to which movements can be predicted only from the planning CT and structures. As with other classical and deep learning registration algorithms, DAM would struggle to register rectum structures due to the randomness in their intensity values. Following clinical practice, we opted for masking the rectum voxels with air. As a result, all deformed CTs have air-filled rectum structures, which can affect the accuracy in the dose calculation, especially for beams delivered in the anterior-posterior direction. Possible solutions include adding an additional generative model that generates rectum voxel intensities based on the organ mask shape.

\section{Conclusion}
\label{sec:Conclusion}
We presented DAM, a deep learning-based daily anatomy model to simulate patient-specific deformations that may be observed during the course of a prostate cancer radiotherapy treatment. DAM captures groups of correlated movements via few auxiliary latent variables, where few variables are able to model prostate deformations and shifts with similar accuracy as state-of-the-art models based on principal component analysis. Compared to previous population models, DAM can generate realistic CT images and contours in less than a second without any pre-processing, with volume changes and center of mass shifts that match in frequency and range those reported in the clinics and in previous studies. Given its simplicity and speed to generate CTs based on a single planning scan and delineations, DAM can be tested in treatment planning and evaluation to design treatment plans that are robust against inter-fraction variations.

\section*{Acknowledgments}
This work is supported by KWF Kanker Bestrijding [grant number 11711] and is part of the KWF research project PAREL. Zolt\'an Perk\'o would like to thank the support of the NWO VENI grant ALLEGRO (016.Veni.198.055) during the time of this study. We would like to thank Haukeland University Hospital (Bergen, Norway), responsible oncologist Svein Inge Helle, physicist Liv Bolstad Hysing and Dr. Jeffrey Siebers for providing the CT-data with contours. 

\section*{Code availability}
The code, weights and results are publicly available at \url{https://github.com/}.

\section*{CRediT authorship contribution statement}
\textbf{Oscar Pastor-Serrano}: Conceptualization, Data Curation, Methodology, Investigation, Formal Analysis, Visualization, Software, Writing - original draft, Writing - review \& editing. \textbf{Steven Habraken}: Resources, Formal Analysis, Writing - review \& editing. \textbf{Mischa Hoogeman}: Conceptualization, Formal Analysis, Project administration, Funding acquisition, Writing - review \& editing. \textbf{Danny Lathouwers}: Formal Analysis, Writing - review \& editing. \textbf{Dennis Schaart}: Formal Analysis, Writing - review \& editing. \textbf{Yusuke Nomura}: Formal Analysis, Writing - review \& editing. \textbf{Lei Xing}: Supervision, Formal Analysis, Writing - review \& editing. \textbf{Zoltán Perkó}: Conceptualization, Supervision, Methodology, Formal Analysis, Funding acquisition, Project administration, Writing - review \& editing.

\small
\bibliographystyle{bibstyle}
\bibliography{organ_models}

\normalsize
\appendix
\section{Diffeomorphic transformations}
\label{App:scsq}
% deriving elbo
In this section, we provide details about the type of diffeomorphic transformation used in our model, based on the seminal works in \cite{ashburner_fast_2007, dalca_unsupervised_2019}. The chosen diffeomorphic transformation is represented via the ordinary differential equation 

\begin{equation}
	\frac{\partial\Phi^{(t)}}{\partial t} = v(\Phi^{(t)})
\end{equation}

\noindent describing the evolution of the deformation over time, where $t\in[0,1]$ is time, $\Phi^{(0)}$ is the identity transformation and $v:\mathbb{R}^3\rightarrow\mathbb{R}^3$ is the stationary velocity field. To generate a DVF, we start from the identity transformation $\Phi^{(0)}$, integrating over time to obtain $\Phi^{(1)}$. In our case, we scaling and squaring \cite{moler_nineteen_2003, arsigny_log-euclidean_2006}, which involves recursively updating the DVF in $T$ successive small time steps

\begin{equation}
	\Phi^{(1/2^T)}=\bm{p}+v(\bm{p})/2^T
\end{equation}

\begin{equation}
	\Phi^{(1/2^{t-1})}=\Phi^{(1/2^t)}\circ\Phi^{(1/2^t)}
\end{equation}

\begin{equation}
	\Phi^{(1)}=\Phi^{(1/2)}\circ\Phi^{(1/2)}
\end{equation}

\noindent where $\bm{p}$ are spatial locations. Typically, $T$ is chosen so that $v(\bm{p})/2^T$ is small, with higher $T$ leading to more accurate solutions. In Group theory, the velocity field $\bm{v}$ is a member of the Lie algebra, which is exponentiated to produce the member of the Lie group $\Phi^{(1)} = \exp{v}$, establishing the connection between the exponentiation and the integration of the ordinary differential equation. %The integration of the ordinary differential equation represents a one-parameter subgroup of diffeomorphisms, which, for scalars $t_i$ and $t_j$ and the composition $\circ$ associated with the Lie group, implies that $\exp{(t_i+t_j)}v = \exp{t_iv}\circ\exp{t_jv}$, based on the properties of one-parameter subgroups.

\section{Evidence lower bound}
\label{App:elbo}
The lower bound (LB) derivation is bassed on Jensen's inequality. For concave functions such as the natural logarithm and a random variable $x$, Jensen's inequality states that 

\begin{equation}
	\log\:\big(\mathbb{E}[x]\big)\geq\mathbb{E}\:[\log(x)].
\end{equation}

Starting from the marginal likelihood of the probabilistic model in \egyref{marg}, the lower bound is obtained as

\begin{align}
	\begin{split}
		\label{eq:1}
		\log\:(P_{\bm{\theta}}(\bm{y}|\bm{x},\bm{s}_x)) ={}& \log\: \int P_{\bm{\theta}}(\bm{y}|\bm{z},\bm{x},\bm{s}_x)P(\bm{z}|\bm{x},\bm{s}_x)d\bm{z}
	\end{split}\\
	\begin{split}
		\label{eq:2}
		={}& \log\: \int P_{\bm{\theta}}(\bm{y}|\bm{z},\bm{x},\bm{s}_x)P(\bm{z}|\bm{x},\bm{s}_x) \frac{Q_{\bm{\phi}}(\bm{z}|\bm{x},\bm{s}_x,\bm{y},\bm{s_y})}{Q_{\bm{\phi}}(\bm{z}|\bm{x},\bm{s}_x,\bm{y},\bm{s_y})}d\bm{z}
	\end{split}\\
	\begin{split}
		\label{eq:3}
		={}& \log\: \mathbb{E}_{\mathrm{\mathbf{z}}\sim Q_{\bm{\phi}}(\bm{z}|\bm{x},\bm{s}_x,\bm{y},\bm{s_y})} \Big[\frac{P_{\bm{\theta}}(\bm{y}|\bm{z},\bm{x},\bm{s}_x)P(\bm{z}|\bm{x},\bm{s}_x)}{Q_{\bm{\phi}}(\bm{z}|\bm{x},\bm{s}_x,\bm{y},\bm{s_y})}\Big]
	\end{split}\\
	\begin{split}
		\label{eq:4}
		\geq{}& \mathbb{E}_{\mathrm{\mathbf{z}}\sim Q_{\bm{\phi}}(\bm{z}|\bm{x},\bm{s}_x,\bm{y},\bm{s_y})} \Big[\log\:\Big(\frac{P_{\bm{\theta}}(\bm{y}|\bm{z},\bm{x},\bm{s}_x)P(\bm{z}|\bm{x},\bm{s}_x)}{Q_{\bm{\phi}}(\bm{z}|\bm{x},\bm{s}_x,\bm{y},\bm{s_y})}\Big)\Big]
	\end{split}\\
	\begin{split}
		\label{eq:6}
		={}& \mathbb{E}_{\mathrm{\mathbf{z}}\sim Q_{\bm{\phi}}(\bm{z}|\bm{x},\bm{s}_x,\bm{y},\bm{s_y})} [\log\:P_{\bm{\theta}}(\bm{y}|\bm{z},\bm{x},\bm{s}_x)] - D_{KL}(Q_{\bm{\phi}}(\bm{z}|\bm{x},\bm{s}_x,\bm{y},\bm{s_y})||P(\bm{z}|\bm{x},\bm{s}_x)),
	\end{split}
\end{align}

\noindent where the KL-divergence $D_{KL}$ is defined as 

\begin{equation}
	D_{KL}(P(x)||Q(x)) = \int\log\Big(\frac{P(x)}{Q(x)}\Big)\:P(x)\:dx = \mathbb{E}_{\text{x}\sim P(x)}\log\Big(\frac{P(x)}{Q(x)}\Big).
	\label{eq:kl}
\end{equation}

\end{document}